\documentclass[conference]{IEEEtran}
\pdfoutput=1
\usepackage{mathtools}
\usepackage{newtxmath}
\usepackage{IEEEtrantools}
\usepackage[noadjust]{cite}
\usepackage{algorithm}
\usepackage[utf8]{inputenc}

\newcommand{\mertstitle}{Multi-functional Coexistence of Radar-Sensing and Communication Waveforms}
\DeclareMathOperator*{\argmin}{\arg\,\min}

\newcommand{\diag}[1]{\operatorname{diag}\left({#1}\right)}

\makeatletter
\def\IEEElabelanchoreqn#1{\bgroup
\def\@currentlabel{\p@equation\theequation}\relax
\def\@currentHref{\@IEEEtheHrefequation}\label{#1}\relax
\Hy@raisedlink{\hyper@anchorstart{\@currentHref}}\relax
\Hy@raisedlink{\hyper@anchorend}\egroup}
\makeatother
\newcommand{\subnumberinglabel}[1]{\IEEEyesnumber \IEEEyessubnumber*\IEEElabelanchoreqn{#1}}

\newcommand*{\matr}[1]{\mathbf{#1}}

\usepackage{dblfloatfix}
\usepackage{balance}
\usepackage{subfig}

\usepackage{prettyref}
\newrefformat{fig}{Fig.~\ref{#1}}
\newrefformat{conj}{Conjecture~\ref{#1}}
\newrefformat{tab}{Table~\ref{#1}}
\newrefformat{sec}{Section~\ref{#1}}
\newrefformat{subsec}{Section~\ref{#1}}
\newrefformat{subsubsec}{Section~\ref{#1}}
\newrefformat{algo}{Algorithm~\ref{#1}}

\usepackage{amsmath,amssymb,amsfonts}
\usepackage{algorithmic}
\usepackage{graphicx}
\usepackage{textcomp}
\usepackage{xcolor}

\usepackage{acronym}
\usepackage{textcomp}

\usepackage{siunitx}
\DeclareSIUnit{\belmilliwatt}{Bm}
\DeclareSIUnit{\dBm}{\deci\belmilliwatt}
\DeclareSIQualifier{\isotropic}{i}
\DeclareSIQualifier{\carrier}{c}

\makeatother

\usepackage{listings}
\def\BibTeX{{\rm B\kern-.05em{\sc i\kern-.025em b}\kern-.08em
		T\kern-.1667em\lower.7ex\hbox{E}\kern-.125emX}}
\begin{document}	
	\title{\mertstitle}

	\author{Mehmet Mert Şahin\IEEEauthorrefmark{1}, Hüseyin Arslan\IEEEauthorrefmark{1}\IEEEauthorrefmark{2}\IEEEmembership{, Fellow, IEEE}\\\IEEEauthorblockA{\IEEEauthorrefmark{1}Department of Electrical Engineering, University of South Florida, Tampa, FL, 33620}\IEEEauthorblockA{\IEEEauthorrefmark{2}Department of Electrical and Electronics Engineering, Istanbul Medipol University, Istanbul, TURKEY, 34810}e-mail: mehmetmert@usf.edu, arslan@usf.edu}
	
\maketitle
\thispagestyle{plain}
\pagestyle{plain}

\begin{abstract}
 In this study, a novel transmission scheme is proposed to serve radar-sensing and communication objectives at the same time and allocated bandwidth. The proposed transmitted frame non-orthogonally superimposes two different waveforms, which are frequency modulated continuous-wave (FMCW) for radar-sensing and orthogonal frequency division multiplexing
(OFDM) for communication. Also, the receiver scheme that performs channel estimation via radar-sensing functionality without degrading data rate of communication operation is introduced. As numerically evaluated, the proposed system achieves good sensing accuracy even if the signal-to-noise ratio (SNR) is low, and communication performance is only \SI{0.6}{\deci\bel} less at the target bit-error rate (BER) of 1\% compared to the assumption of perfect channel state information (CSI) without any pilot overhead over OFDM subcarriers.  
 
    \acresetall
\end{abstract}	
	
	\section{Introduction}	
Nowadays, evolved hardware such as antenna arrays, and digital signal processing (DSP) chips; and software regarding algorithms for detection and estimation capabilities lead radar and communication systems to intersect in order to provide efficient usage of resources. Making these two different worlds to work in harmony may urge lots of promising applications that pave the way for new techniques in wireless technologies. Therefore, this trend attracts lots of interest from both industry and academia \cite{hanzo_2020}.   




With the usage of millimeter wave (mmWave) bands, directed beams and larger bandwidths are served that improve sensing accuracy and data rate. Most of papers pointing convergence of communication and radar-sensing investigate optimal waveform, which is called joint radar-communication
(JRC) waveform, to serve both two applications properly \cite{HeathVorobyov2020}. The aim is to combine radar-sensing and communication into a single mmWave system that utilizes a standard waveform. This kind of system is aimed to be optimized regarding cost, size, power consumption, spectrum usage, and adoption of communication-capable vehicles, for example, in case of autonomous driving, which needs both radar and vehicle-to-vehicle (V2V) communication \cite{bliss_2017}. 



One of the JRC waveform design has been studied in \cite{Sturm_2011}, where the system associates part of
OFDM or spread spectrum resources for radar processing by transmitting known sequences using these waveforms. The drawback of this scheme 
is the lack of degrees of freedom leading to multiplexing
of data between known sequence assigned for radar and unknown information sequence assigned for communication. Orthogonal multiplexing or modification of one waveform serving both applications prevent the full exploitation of the time-frequency-space resources in terms of spectral efficiency and radar-sensing performance. Also, OFDM-radar systems lack high resolutions compared to the FMCW radars that are mostly used in automotive radar \cite{Ali_2017}. 
	
Besides, some JRC waveforms modulate communication messages on top of the radar waveforms, using phase modulation techniques in \cite{Metcalf_2019, Ottersten2019}. However, compared to plain OFDM modulation, these kinds of modulation schemes degrades spectral efficiency. Moreover, the JRC
waveforms are proposed by separating communication information via the transmit beamforming vectors \cite{ahmad_2016}. Similarly, these methods suffer from low data rates since the communication signal must be spread to avoid interference over the radar-sensing functionality. A RadChat unit proposed in \cite{BlissWymeersch2019} includes two different waveforms for radar and communication use cases, which are FMCW and narrowband transmission, respectively. However, the system does not transmit both waveforms at the same time, the transmission is switched between radar and communication. Reference \cite{ReedSpooner2019} studies the interference of linear-frequency-modulated radar on OFDM signal where two different use-cases, communication and radar, are considered. However, the transmission is not coordinated, and only the interference issue is studied.  


\begin{figure}
  \centerline{\includegraphics[width=1\linewidth]{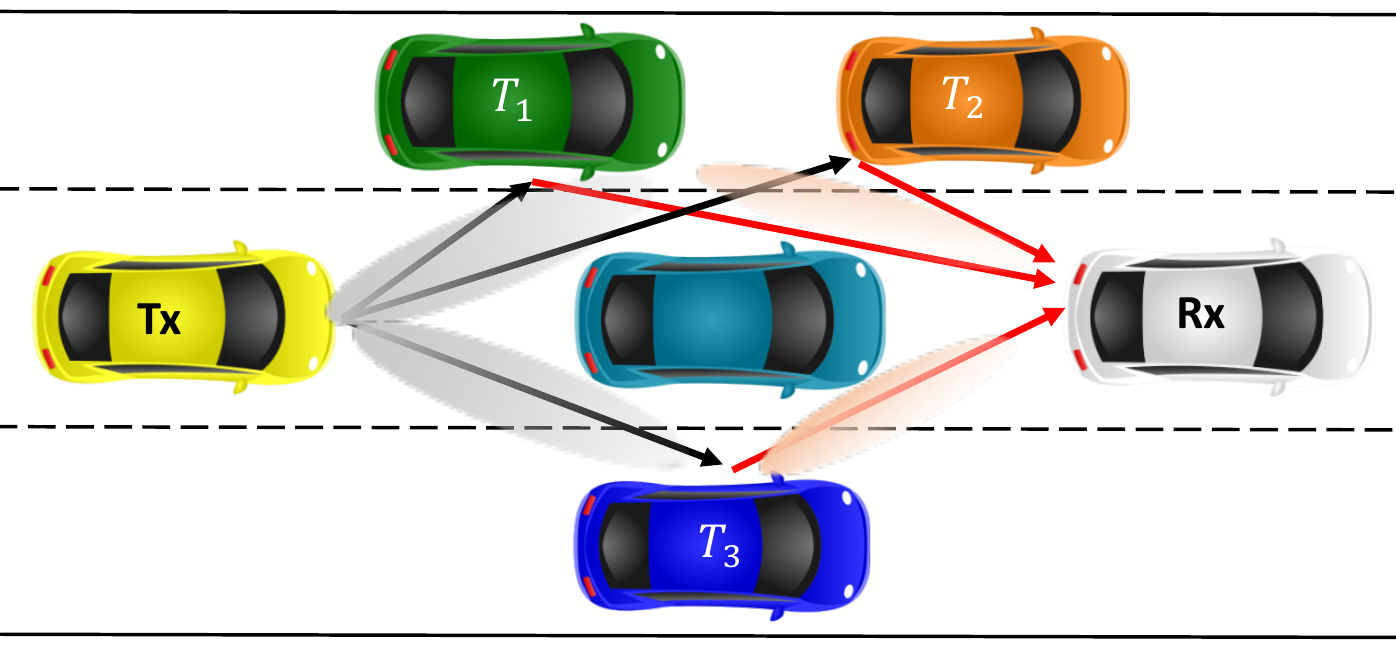}}
\caption{System model needing both radar-sensing and communication functionality.}
\label{fig:systemModel}
\end{figure}	

In this study, the proposed receiver is assigned as  a bi-static radar-sensing and communication receiver where the transmitter and receiver are not co-located like in \prettyref{fig:systemModel}. Although many of the previous studies have addressed JRC assuming vehicles equipped with mono-static radars supporting communication capability \cite{ottersen2018ofdm_based}, the bi-static radar system which provides radar-sensing information to the receiver can be used to improve communication functionality. We propose a novel scheme that non-orthogonally overlaps two separate waveforms (OFDM and FMCW) peacefully serving two different applications, as suggested in our previous work on waveform-domain non-orthogonal multiple access (NOMA) investigating the coexistence of two different waveforms for communication functionality  \cite{sahin_2020waveformdomain}. Here, we also perform channel estimation via the FMCW waveform, which is offered for radar-sensing to demodulate communication symbols in OFDM without using any pilot overhead.  



	
\section{System Model} \label{sec:SystemModel}

\begin{figure*}
  \includegraphics[width=\textwidth]{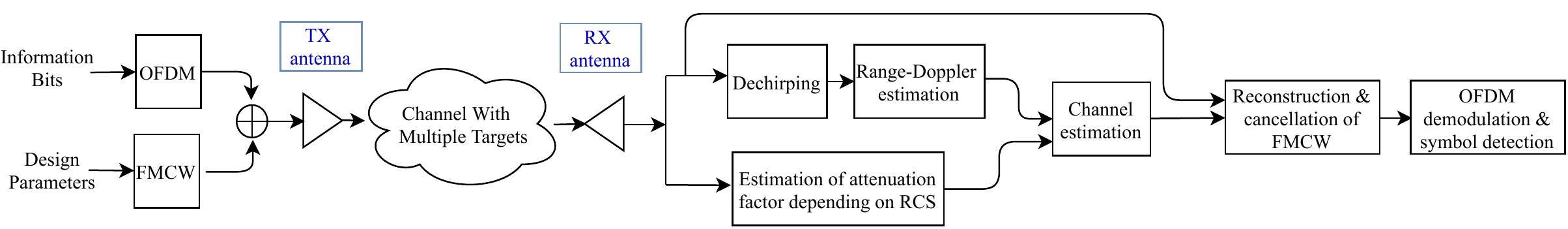}
  \caption{The transmit and receive diagram of the proposed scheme.}
  \label{fig:generalFrame}
\end{figure*}

The V2V scenario is considered as shown in \prettyref{fig:systemModel}, however, the proposed scheme is also applicable for different kinds of use cases needing both radar-sensing and communication activity. The general transmission and reception scheme can be seen in \prettyref{fig:generalFrame}. 

\subsection{Transmission Design}

FMCW consisting of many chirps and OFDM waveforms are utilized for radar-sensing and communication operations, respectively. 	
The complex equivalent time-domain representation of one chirp sweeping linearly across a total bandwidth of $\beta$ Hz during the $\tau$-second is expressed as \cite{richards_2014}
\begin{equation} \label{eq:chirp}
s_\text{chirp}(t) = e^{j\pi\beta t^2/\tau}, \qquad   0\leq t\leq \tau.
\end{equation}
The FMCW waveform consisting of K chirps per frame is
\begin{equation} \label{eq:fmcw}
s_\text{FMCW}(t) = \sqrt{P_\text{FMCW}} \sum_{k=0}^K s_\text{chirp}(t-k\tau), \qquad   0\leq t\leq T,
\end{equation}
where $T$ and $P_\text{FMCW}$ denote the total duration and power of FMCW waveform, respectively.

Denote $\{d_n\}_{n=0}^{N-1}$ as the complex symbols modulated via quadrature amplitude modulation (QAM) drawn from a complex symbol alphabet $\mathbb{S}$. The OFDM signal in time-domain is expressed as 
\begin{equation} \label{eq:ofdm}
s_\text{OFDM}(t) = \sqrt{P_\text{ofdm}} \sum_{n=0}^{N-1} d_n e^{j 2\pi n \Delta f t}, \qquad   0\leq t\leq T_s,
\end{equation}
where $T_s$, $\Delta f$ and $P_{\text{OFDM}}$ denote one OFDM symbol duration, the subcarrier spacing and the power of OFDM, respectively.

A cyclic prefix (CP) of length $T_g$ is prepended to each  OFDM symbol to keep OFDM subcarriers orthogonal by preventing intersymbol interference (ISI) across OFDM symbols and transform the linear convolution of the multipath channel to a circular convolution which eases equalization \cite{stuber_2011}. After the CP addition, the $m$th OFDM symbol can be expressed as
\begin{equation} \label{eq:ofdmCP}
\Bar{s}_m(t) = \left\{ \,
\begin{IEEEeqnarraybox}[][c]{l?s}
\IEEEstrut
s_{\text{OFDM}}(t+T_s-T_g) & if $0\leq t\leq T_g$, \\
s_{\text{OFDM}}(t-T_g) & if $T_g < t \leq T_{\text{OFDM}} $,
\IEEEstrut
\end{IEEEeqnarraybox}
\right.
\end{equation}
where $T_{\text{OFDM}} = T_g + T_s$ is the duration of one OFDM symbol after CP addition. 
Having $M$ OFDM symbols in a frame during $T_{\text{sym}} \leq T$, the time domain OFDM signal can be represented as 
\begin{equation} \label{eq:ofdmSymbols}
\Bar{s}_{\text{OFDM}}(t) = \sum_{m=0}^{M-1} \Bar{s}_m(t) \times \text{rect}\left(\frac{t-mT_{\text{OFDM}}}{T_{\text{OFDM}}}\right), \quad 0\leq t\leq T_{\text{sym}}.
\end{equation}

Revisiting the expression of two different waveforms OFDM and FMCW, the transmitted frame $s(t)$ for the objectives of multi-functional radar-sensing and communication transmission is designed as follows: 
\begin{equation} \label{eq:transmittedFrame}
{s}(t) = \left\{ \,
\begin{IEEEeqnarraybox}[][c]{l?s}
\IEEEstrut
s_\text{FMCW}(t) & if $0\leq t\leq \tau$, \\
s_\text{FMCW}(t) + \Bar{s}_{\text{OFDM}}(t) & if $\tau < t \leq T $.
\IEEEstrut
\end{IEEEeqnarraybox}
\right.
\end{equation}

\begin{figure}
\subfloat[]{
\vspace{-0.5cm}
    \centerline{\includegraphics[width=0.9\linewidth]{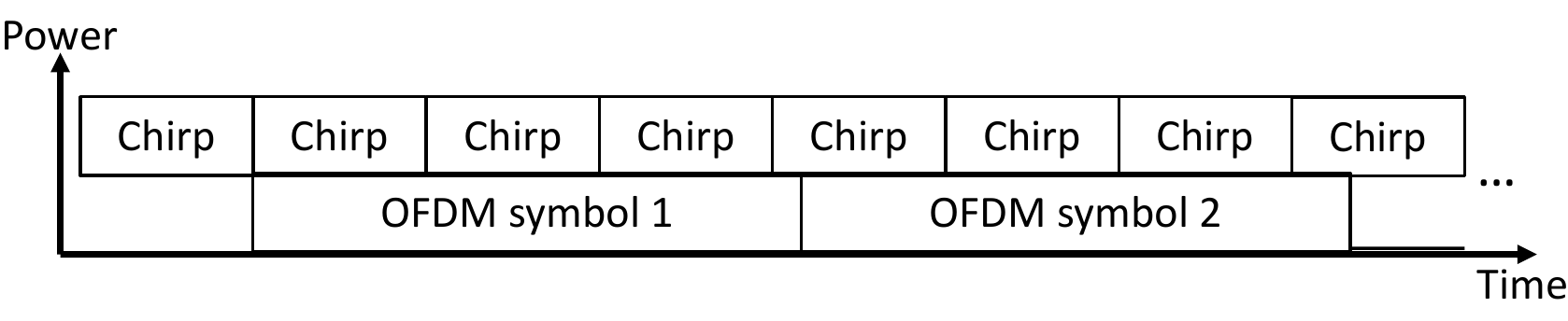}}
    \label{fig:txFrame} }
\newline
\vspace{-0.2cm}
\subfloat[]{
  \centerline{\includegraphics[width=0.9\linewidth]{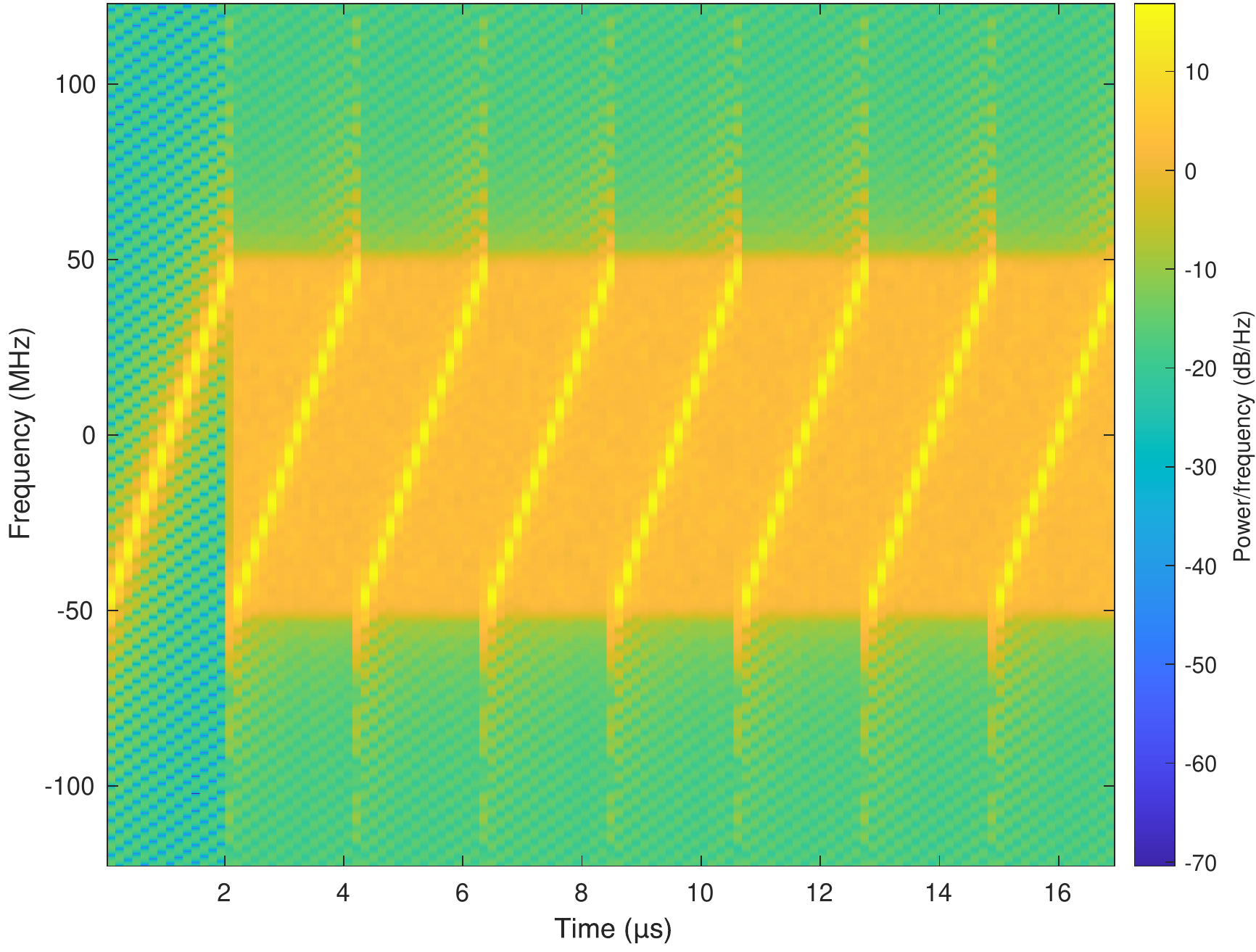}}
\label{fig:txFrameSpectogram}
}
\caption{Transmitted frame design for joint radar and communication functionality, (a) time-power representation, (b) time-frequency representation.}
\vspace{-0.5cm}
\end{figure}

The transmitted frame can be seen in \prettyref{fig:txFrame} where it starts with a chirp following the superimposed OFDM symbols and many chirps. The waveforms are superimposed in a way that the allocated bandwidth is the same for each waveform type. The time-frequency illustration of the superimposed signal can be seen in \prettyref{fig:txFrameSpectogram}. It can be realized that the OFDM signal is distributed along with the whole bandwidth, whereas FMCW waveform patterns consecutive pulses whose frequency increases linearly.    

Then, the baseband signal $s(t)$ is upconverted to the desired radio frequency (RF) band, where the transmitted passband analog signal becomes 
\begin{equation} \label{eq:passBand}
x(t) = \Re{\{s(t)e^{j (2\pi f_c t + \Bar{\theta})}\}},
\end{equation}
where $\Re{\{.\}}$ denotes the real part of the complex quantity. The notations $f_c$ and $\Bar{\theta}$ are the carrier frequency in which most automotive radars operate in \SI{24}{\giga\hertz} or \SI{76}{\giga\hertz} bands \cite{Ali_2017}, and the initial phase of the transmitted signal, respectively.

\subsection{Channel Effect}


The Doppler shift and flight time for the paths reflecting from the targets are as shown in \prettyref{fig:systemModel} and are assumed to be fixed over a coherent transmission time $T$. 
Under the linear time-varying channel, the received passband signal is represented as \cite{hlawatsch_matz_2011}
\begin{IEEEeqnarray}{rCl}
r(t) & = & \sum_{p=1}^P \alpha_p \Re{\{x(t-\tau_p)e^{j2\pi(f_c + \psi_p)(t-\tau_p) + j\Bar{\theta} + j \vartheta_p} \}} + n(t), \IEEEeqnarraynumspace  \label{eq:channelModel} 
\end{IEEEeqnarray}
where $\alpha_p$, $\tau_p$ and $\vartheta_p$ are the attenuation factor depending on nonfluctuating radar cross section (RCS), time delay related with the distance between the transmitter to target plus target
to the receiver (bi-static range) and phase error, respectively. The notation $\psi_p = \frac{f_c \upsilon_p}{c}$ is the Doppler frequency associated with the $p$th path depending on relative speed $\upsilon_p$ and the letter $c$ denotes the speed of light. The number $P$ indicates the number of radar targets; in other words, the number of specular scatterer in the environment. Also, $n(t)\sim\mathcal{CN}\left(0,\sigma^2\right)$ denotes the additive white Gaussian noise (AWGN). The attenuation factor $\alpha_p$ is proportional to the large-scale path-loss. Having the path distance $d$ between receiver and transmitter, the large-scale path-loss $G$ is given as 
\begin{equation} \label{eq:pathloss}
	G = \frac{G_{\text{TX}}G_{\text{RX}}\lambda^2}{(4\pi)^2d^\text{PL}}, 
\end{equation}	
where $\text{PL}$ is the path-loss exponent, $G_{\text{TX}}$ and $G_{\text{RX}}$ are the transmit and receive antenna gain, respectively.


\section{Multi-functional Reception}\label{sec:reception}
In this section, the receiver scheme for radar-sensing and communication operations is investigated. Since the receiver performs both functions, the knowledge obtained from one process can be leveraged to another to improve the performance.  

\subsection{Bi-static Radar Functionality and Channel Estimation}

Down-converting the received passband signal $r(t)$ into baseband and sampling with the frequency of $F_s = N\Delta f$, the discrete-time signal becomes 
\begin{IEEEeqnarray}{rCl} 
\subnumberinglabel{eq:block}
y[n] & = & \sum_{p=1}^P h_p x \left( n/F_s-\tau_p \right) e^{j2\pi n \psi_p / F_s} + w(n), \,\, n \in \mathbb{N}^+, \IEEEeqnarraynumspace  \label{eq:receivedSampled} \\
\noalign{\noindent and
\vspace{2\jot}}
h_p & = & \alpha_p e^{-j 2\pi (f_c + \psi_p)\tau_p + j\Bar{\theta}+j\vartheta_p}, \IEEEeqnarraynumspace  \label{eq:receivedCoeff}
\end{IEEEeqnarray}
where $h_p$ is the complex channel gain of target $p$.
Then, the stretch processing is employed in the discrete domain for the superimposed signal to get delays and Doppler shifts estimations. The processed signal in one chirp time interval can be written as
\begin{equation} \label{eq:chirp}
\Bar{y}[n^\prime] = y[n^\prime] \times e^{-j\pi\beta (n^\prime/F_s)^2/\tau}, \qquad n^\prime = 1,2,\ldots, \tau F_s,
\end{equation}
and dechirping process is repeated for each chirp time interval. Remember that stretch processing is generally done in time domain with down-conversion. However, here it is assumed that the occupied bandwidth of FMCW is the same as OFDM bandwidth, therefore, the sampling rate $F_s$ for both radar and communication is taken as equal to each other.   
 
The first step is to form a fast-time/slow-time coherent processing interval (CPI) matrix. Fast time samples are obtained at the rate of $F_s$ from the points on each chirp, which are called range bins. Slow-time samples are taken from the points on every chirp at the same range bins, which are called Doppler bins. The range-Doppler matrix is obtained by 2D-FFT over CPI matrix yielding peaks at the intersection of $\tau_p$ and $\psi_p$ for each $p$th target. 
Well-known algorithms such as constant-false alarm rate (CFAR) processing and space-time adaptive processing (STAP) can be used to determine the threshold for the target detection properly \cite{Ali_2017}. However, the investigation of these algorithms is not the scope of this paper. The threshold is determined after several Monte-Carlo simulations. 

Estimation of complex attenuation factor $h_p$ for every  $p$th scatterer (target) is done in the first chirp with the least-square estimation \cite{arslan_bottomley_2001}. It is worth to stress that first chirp in the transmitted frame is interference-free as seen in \prettyref{fig:txFrame}. Let the vector $\matr{y_c} \triangleq [y[1], y[2],\ldots,y[\tau F_s]]^\text{T}$ be the samples of the received signal throughout the time $\tau$, the estimated complex attenuation coefficients $\matr{\hat{h}} \triangleq [\hat{h}[1], \hat{h}[2],\ldots,\hat{h}[P]]^\text{T}$ become
\begin{equation} \label{eq:LScoeff}
\matr{\hat{h}} = \argmin_{\matr{h}} (\matr{y_c}-\matr{B}\matr{h})^H(\matr{y_c}-\matr{B}\matr{h}),
\end{equation}
where $\matr{B}$ is a $(\tau F_s - \Bar{n})\times P$ matrix whose rows corresponds to different shifts of the transmitted chirp where shift values are determined according to estimated range value of the $p$th scatterer (target). The offset value $\Bar{n} \in \mathbb{N}$ depends on the maximum range requirement of the system. Also, the selection of higher value for $\Bar{n}$ decreases fluctuations in the estimation of $\matr{\hat{h}}$ depending on Doppler shifts along with one chirp, whereas the maximum range is reduced.    
By differentiating with respect to $\matr{h}$ and setting the result equal to zero, the least-square estimation of the channel becomes 
\begin{equation} \label{eq:LScoeff}
\matr{\hat{h}} = (\matr{B}^H\matr{B})^{-1}\matr{B}^H\matr{y}.
\end{equation}
Besides the estimation of delays $\tau_p$ and Doppler shifts $\psi_p$, the matrix $\hat{\matr{h}}$ completes the process to recreate the channel matrix $\matr{H}$ with some estimation errors.

\subsection{Communication Functionality}
Here, the communication symbols are decoded using the estimated channel in the previous section. Let the channel gain of the $k$th sample of the transmitted signal during the reception of the $n$th sample denote as $h_{n,k}$. Also, if the discrete channel convolution matrix along one OFDM symbol duration with $N_{\text{OFDM}}$ samples is shown as $\matr{H} \in \mathbb{C}^{(N_{\text{OFDM}})\times(N_{\text{OFDM}})}$, the element in the $k$th column of the $n$th row of $\matr{H}$ is $h_{n,k}$. 
Firstly, the FMCW sequence is removed from the total received signal by using estimated channel matrix $\matr{\hat{H}}$ as follows: 
\begin{equation} \label{eq:LScoeff}
\matr{y_{\text{OFDM}}} = \matr{y}-\matr{\hat{H}}\matr{s_{\text{FMCW}}},
\end{equation}
where $\matr{y_{\text{OFDM}}}= \left[y_1, y_2, \ldots, y_M\right]$ and $y_m$ is the $m$th OFDM symbol in the received vector $y$. 
The CP addition matrix $\matr{A} \in \mathbb{R}^{N_{\text{OFDM}} \times N}$ is defined as 
\begin{equation}\label{eq:cpAddition}
    A=\left[
\begin{array}{c} 
    \begin{array}{cc}
\matr{0}_{N_g\times N} & \matr{I}_{N_g}
\end{array}
\\  \matr{I_{N_{\text{OFDM}}}}
\end{array}
\right],
\end{equation} 
and the CP removal matrix is generated as $B = \left[\matr{0}_{N\times N_g} \matr{I}_N \right]$ where $N_g$ is the total sample number during CP duration $T_g$.
\begin{figure} 
  \centerline{\includegraphics[width=1\linewidth]{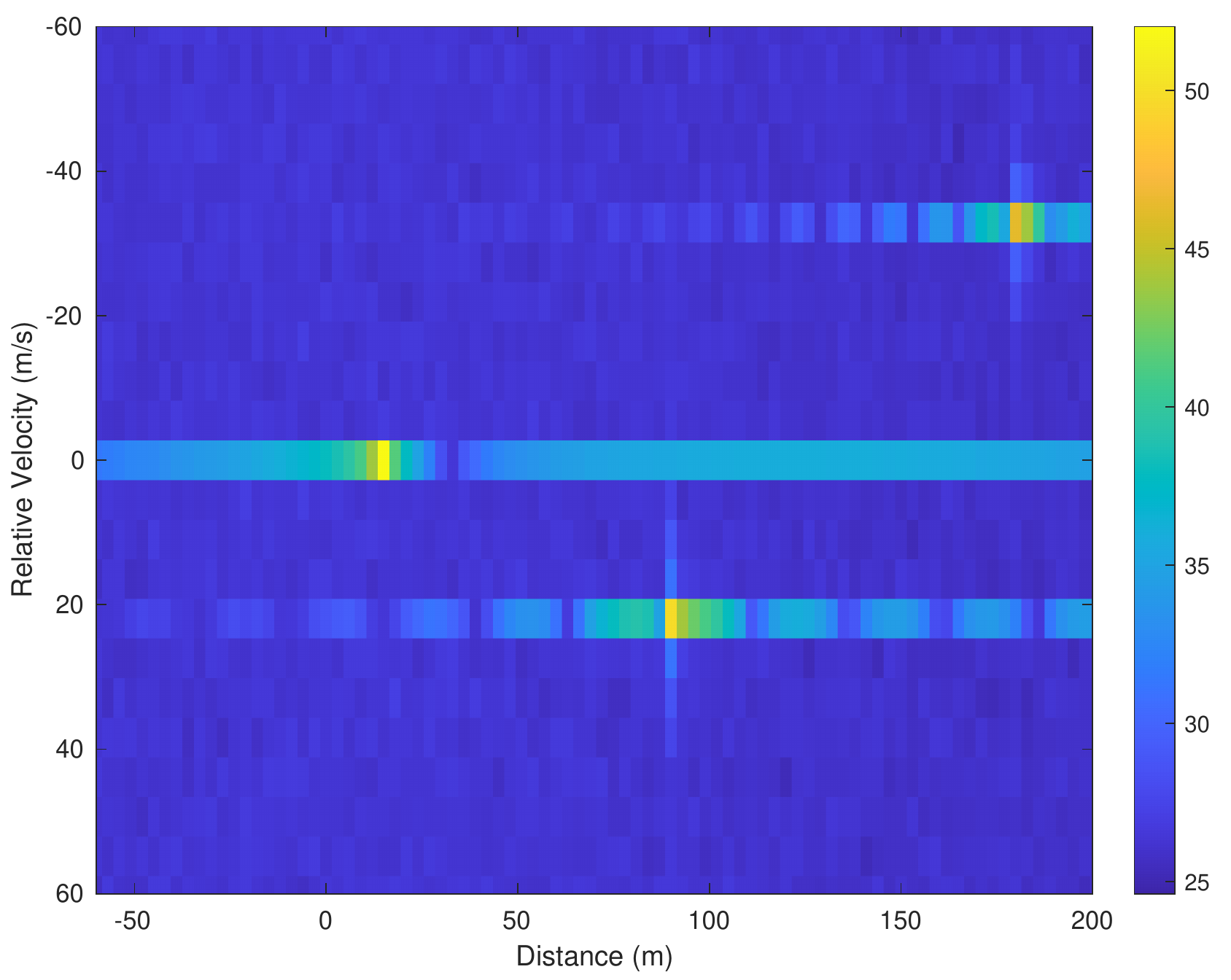}}
\caption{Distance-velocity plot of the targets where SNR equals to $\SI{20}{\deci\bel}$.}
\label{fig:delayDoppler}
\vspace{-0.5cm}
\end{figure}
The matrix $\matr{\Theta} \in \mathbb{C}^{N\times N}$ is the complete channel frequency response (CFR) matrix which equals to 
\begin{equation}
    \matr{\Theta} = \matr{F}_N \matr{B} \matr{\hat{H}} \matr{A} \matr{F}_N^{\text{H}}.
    \label{eq:channel}
\end{equation}
The diagonal components of \prettyref{eq:channel} are the channel coefficients scaling the subcarrier in interest collected in a vector $\boldsymbol{\theta} = \diag{\matr{\Theta}}$ and off-diagonal components are not zero due to Doppler effect from the channel causing ISI. Finally, estimates of data symbols consisting of information bits are obtained as: 
\begin{equation}
    \matr{\hat{d}_m} = \left( \left( \diag{\boldsymbol{\theta} \odot \boldsymbol{\theta}^*} \right)^{-1} (\diag{\theta}^*) \matr{F}_N \matr{B}_K \matr{y}_m \right).
\end{equation}
This equation finalizes the proposed receiver structure without introducing pilots on the OFDM subcarriers to estimate the channel.
\vspace{-0.22cm}
\section{Numerical Results}
\vspace{-0.1cm}
\begin{table}[!t]
	\renewcommand{\arraystretch}{1}
	\caption{Simulation Parameters}
	\label{tab:simulationParameters}
	\centering
	\begin{tabular}{c||c}
		\hline
		\bfseries Parameter &  \bfseries Value \\
		\hline\hline
		Carrier frequency ($f_c$) & \SI{28}{\giga\hertz}\\
		\hline
		Bandwidth ($\beta$) & \SI{100}{\mega\hertz}\\
		\hline
		Sample Rate ($F_s$) & \SI{122.88}{\mega\hertz}\\
		\hline
		Chirp duration ($\tau$) & \SI{2.4}{\micro\second}\\
		\hline
		FMCW duration ($T$) & \SI{2}{\milli\second}\\
		\hline
		Subcarrier spacing($\Delta f$) & \SI{60}{\kilo\hertz}\\
		\hline
		Number of FFT($N$) & 2048 \\
		\hline
		Number of allocated subcarriers & 1666 \\
		\hline
		Range of targets &  \SIlist{15;90;180}{\metre}\\
		\hline
		Relative velocity of targets & \SIlist{0;22;-33}{\metre\per\second}\\
		\hline
	\end{tabular}
	\vspace{-0.5cm}
\end{table}
In this section, the radar-sensing and communication performance of the proposed scheme is demonstrated. Simulation parameters depending on radar and communication requirements are shown in \prettyref{tab:simulationParameters}. It is assumed that the maximum delay $\tau$ caused by targets is smaller than the CP length of each OFDM symbol. Even if the bandwidth of FMCW is set as equal to the bandwidth of OFDM, it can be higher to increase radar-sensing resolution and serve multi-user in communication. It is assumed that the channel includes 3 targets, as shown in \prettyref{fig:systemModel}. The power delay profile (PDP) of the channel is determined as an exponentially decaying function where the power of channel coefficient is set as $\vert h_p(\gamma) \vert^2 = \eta e^{-\gamma p}$. Let $\eta$ be the normalization factor, $p$ be the target index, $\gamma = 1$ be the decaying factor; and the amplitude of each tap follows Rayleigh distribution. During the simulation, powers of waveforms $P_\text{OFDM}$ and $P_\text{FMCW}$ equal to each other.

In \prettyref{fig:delayDoppler}, the radar-sensing performance is demonstrated when the signal-to-noise ratio (SNR) is  ($\SI{20}{\deci\bel}$). Velocity and range of three different targets can be estimated with a proper threshold level. 
After evaluating the complex attenuation factor $h_p$ denoted in \prettyref{eq:receivedCoeff}, the estimation of channel matrix $\matr{\hat{H}}$ is created by using the obtained values of Doppler shifts and delays which is done previously via FMCW waveform. Finally, this channel estimation is used to demodulate communication symbols in the OFDM waveform. 

\begin{figure}
\subfloat[]{
    \centerline{\includegraphics[width=0.8\linewidth]{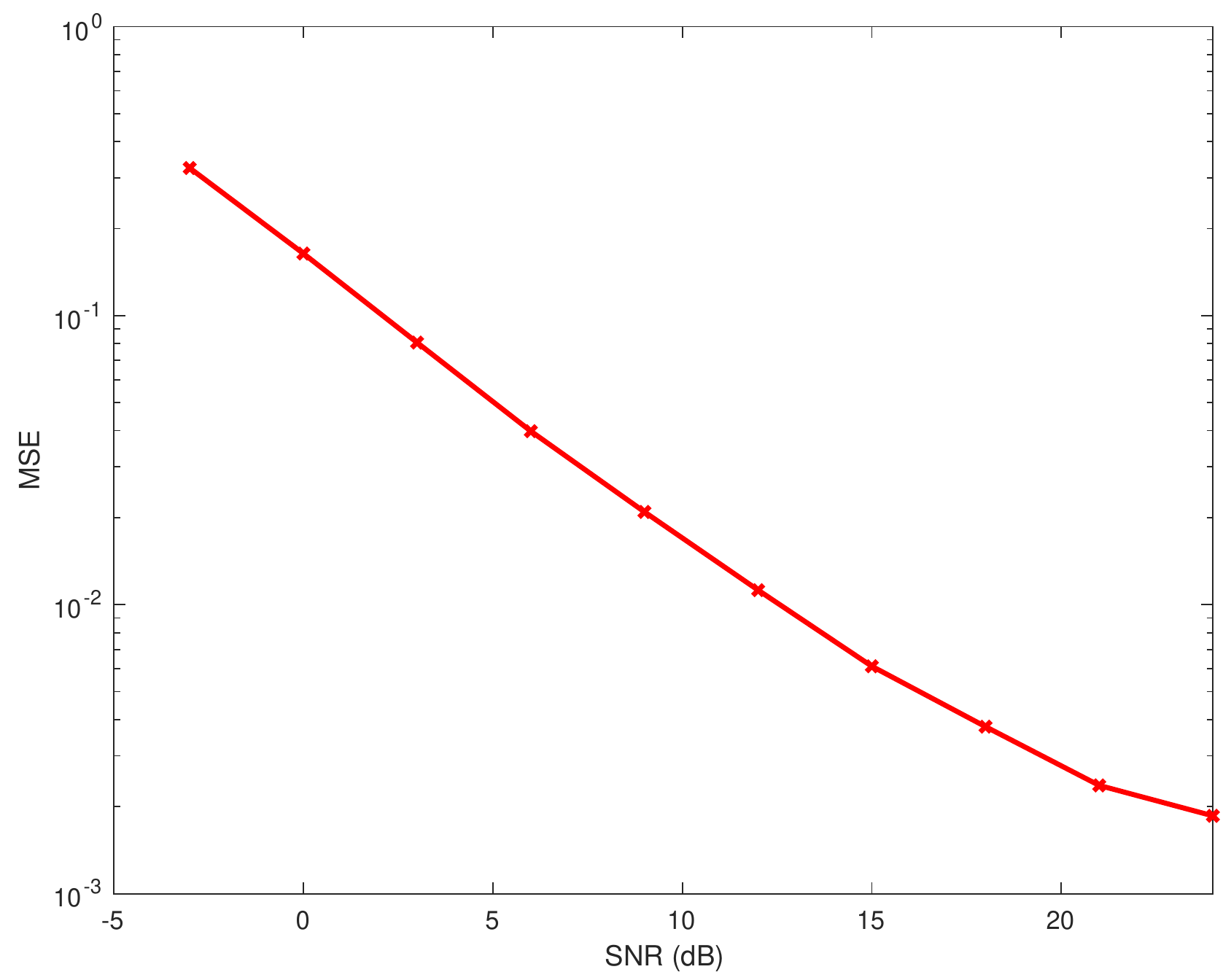}}
    \label{fig:mse}
}
\vspace{-0.3cm}
\newline
\subfloat[]{
  \centerline{\includegraphics[width=0.8\linewidth]{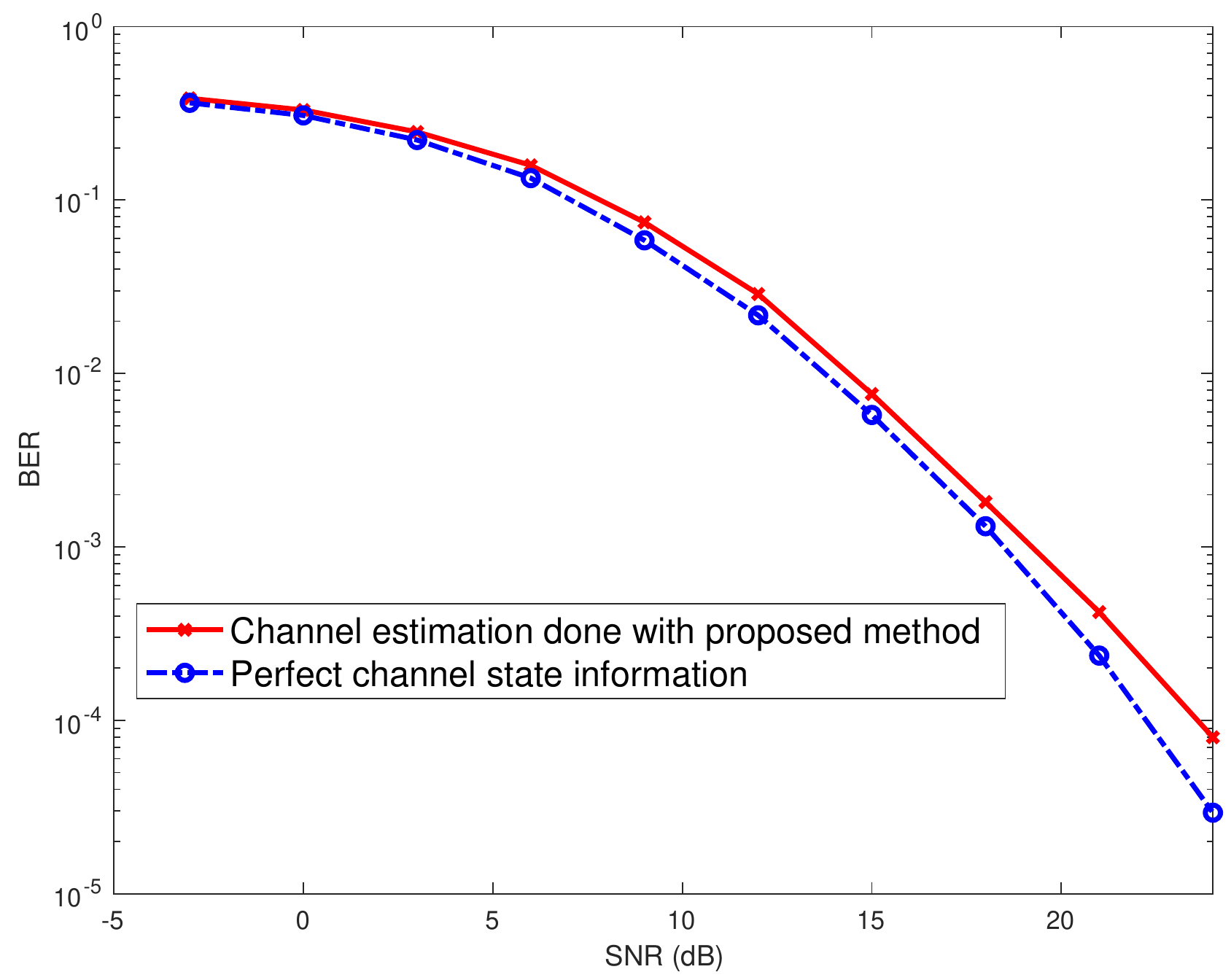}}
\label{fig:berResult}
}
\caption{Performance evaluation of proposed method (a) MSE of channel matrix estimation (b) BER  performance of proposed method.}
\vspace{-0.6cm}
\end{figure}

Mean-square error (MSE) is calculated considering non-zero elements of $\matr{\hat{H}}$ and $\matr{H}$ matrices,  $\sigma_e^2 = \frac{\mathbb{E}[\mid\matr{\hat{H}}-\matr{H}\mid^2]}{\mathbb{E}[\mid \matr{H} \mid^2]}$, where $\mathbb{E[\cdot]}$ denotes the expected value.
MSE performance of the proposed channel estimation scheme can be seen in \prettyref{fig:mse}, which degrades as SNR increases. It is worth noting that the complex attenuation factors for each target are estimated using the first chirp, while delays and Doppler shifts are estimated using the FMCW waveform. 

The BER performance of the information bits transmitted in the OFDM waveform with channel estimation step done in the FMCW waveform can be seen in \prettyref{fig:berResult}. The information bits are encoded via convolutional codes with interleaving to get rid of the deep fading effect of the channel. The proposed scheme is compared with the presence of perfect CSI, which means the FMCW waveform is totally removed from the superimposed signal without affecting the OFDM waveform. The SNR difference is \SI{0.6}{\deci\bel} at the target BER of 1\%, without requiring any pilot symbols over the OFDM symbols.  
\vspace{-0.2cm}
\section{Acknowledgement}
\vspace{-0.15cm}
This work is supported by the National Science Foundation under Grant ECCS-1609581.
\vspace{-0.2cm}
\section{Conclusions}
\vspace{-0.15cm}
In this paper, a novel joint radar-sensing and communication transmission-reception frame is proposed. This frame offers the peaceful coexistence of two different waveforms, such as OFDM and FMCW. This scheme is coordinated at the transmitter, and channel knowledge obtained from the FMCW waveform is used to demodulate OFDM waveform carrying communication information.  
A promising future research direction is to investigate
multi-user implementation with higher bandwidth allocation that could help to increase the range resolution of radar-sensing functionality. Also, more sophisticated algorithms might be studied to confine the transmitted scheme in the desired band and separate waveforms more efficiently that can pave the way for being a part of the future 6G and radio access standards. 

\bibliographystyle{IEEEtran}
\bibliography{comm_rad}
\vspace{-0.3cm}	
\end{document}